\documentclass[aps,prd,superscriptaddress,notitlepage,floatfix,twocolumn]{revtex4-1}
\usepackage{amsmath}
\usepackage{graphicx}   
\usepackage{dcolumn}    
\usepackage{bm}         
\usepackage{amssymb}    
\usepackage{float}
\usepackage{caption}
\usepackage{subcaption}
\usepackage{verbatim}
\usepackage{color}
\usepackage[usenames,dvipsnames,svgnames,table]{xcolor}
\usepackage[colorlinks=true,
            linkcolor=red,
            urlcolor=blue,
            citecolor=green,
            bookmarks=true,
            bookmarksnumbered=true,
            breaklinks=true,
            pdfpagemode=Fullscreen,
            pdfstartview=FitBH]{hyperref}

\newcommand{\gsec}[1]{{\hypersetup{linkcolor=blue}Sec.~\ref{#1}\hypersetup{linkcolor=red}}}
\newcommand{\gfig}[1]{{\hypersetup{linkcolor=violet}Fig.~\ref{#1}\hypersetup{linkcolor=red}}}

\definecolor{gesfred}{rgb}{1,0,0}

\definecolor{gesfblue}{rgb}{0.08,0.42,0.76}

\definecolor{gesfgreen}{rgb}{0.0,0.65,0.0}

\begin{document}

\hfill KEK-TH-1665, arXiv:1308.6522
\vspace{1cm}

\title{Phenomenological Study of Residual $\mathbb Z^s_2$ and $\overline{\mathbb Z}^s_2$ Symmetries}

\author{Andrew D. Hanlon}
\email{ddhanlon@gmail.com}
\affiliation{Department of Physics and Astronomy, Michigan State University, East Lansing, MI 48824}


\author{Shao-Feng Ge}
\email{gesf02@gmail.com}
\affiliation{KEK Theory Center, Tsukuba, 305-0801, Japan}

\author{Wayne W. Repko}
\email{repko@pa.msu.edu}
\affiliation{Department of Physics and Astronomy, Michigan State University, East Lansing, MI 48824}

\date{\today}

\begin{abstract}
The phenomenological consequences of the residual $\mathbb Z^s_2$ and $\overline{\mathbb Z}^s_2$ symmetries are explored in detail. With a precisely measured value of the reactor angle, these two residual symmetries predict distinct distributions for the Dirac {\tt CP} phase and the atmospheric angle, which leads to the possibility of identifying them at neutrino experiments. For both symmetries, it is possible to resolve the neutrino mass hierarchy in most of the parameter space, and they can be distinguished from one another if the true residual symmetry is $\mathbb Z^s_2$ and the atmospheric angle is non-maximal. These results are obtained using an equally split schedule: a $1.5$-year run of neutrinos and a $1.5$-year run of antineutrinos at NO$\nu$A together with a $2.5$-year run of neutrinos and a $2.5$-year run of antineutrinos at T2K. This schedule can significantly increase and stabilize the sensitivities to the mass hierarchy and the octant of the atmospheric angle with only a moderate compromise to the sensitivity of distinguishing $\mathbb Z^s_2$ and $\overline{\mathbb Z}^s_2$.
\end{abstract}

\maketitle

\section{Introduction}

The reactor angle has been accurately measured in the last two years. The first hint of a nonzero reactor angle came from the T2K experiment \cite{T2K11}, followed by MINOS \cite{MINOS11} and Double CHOOZ \cite{DCHOOZ11}, with a confidence level around $3 \sigma$. The next spring, Daya Bay \cite{Daya1203} and RENO \cite{RENO1204} arrived at conclusive measurements, reaching $5.2 \sigma$ and $4.9 \sigma$, respectively. The significance continued climbing to $7.7 \sigma$ \cite{Daya1210} by October of the same year, resulting in a large reactor angle, $\sin^2 2 \theta_{13} = 0.089 \pm 0.010 \pm 0.005$, and providing a great opportunity for further developments in the field of neutrino physics. 

From the theoretical side, new models are needed to accommodate the large reactor angle, see reviews \cite{review1205,*review1301,*review1304,Smirnov12Kyoto,Smirnov13} and references therein. A direct consequence of a nonzero reactor angle is that $\mu$--$\tau$ symmetry has to be broken. It may appear as a part of the full flavor symmetry that constrains the fundamental Lagrangian, but it has to be broken when neutrinos obtain mass. Similarly, all other broken symmetries are hidden, at least in neutrino oscillation experiments, hence ``do not lead to testable predictions" \cite{Smirnov12Kyoto}. If neutrino mixing is really determined by some symmetry, it has to be a residual symmetry that constrains the neutrino mass matrix and hence directly determines the mixing pattern \cite{Ge1104,*Ge1108}. 

With this picture in mind, two approaches \cite{Smirnov13} can be identified in the search of a model to account for the large reactor angle. First, corrections can be added to drive the reactor angle away from zero and the atmospheric angle away from the maximal value with the extent of the deviations depending on model parameters. Or, a residual symmetry can be used to establish a correlation between mixing parameters that is independent of model parameters \cite{Ge1104,*Ge1108}. This unique correlation can predict the reactor angle that is consistent with current experiments, or it can predict the Dirac {\tt CP} phase to be tested by future experiments. Full flavor symmetry can be reconstructed in a bottom-up way from residual symmetries \cite{Smirnov1204,*Smirnov1212} which serves as a lower energy effective theory. 

From the experimental side, the large reactor angle paves the way to measure the remaining parameters of neutrino oscillations. The mass hierarchy can be measured by medium-baseline reactor experiments, such as JUNO \cite{Juno} and RENO-50 \cite{reno50}, atmospheric neutrino experiments, such as PINGU \cite{pingu} and HyperK \cite{hyperk}, as well as accelerator experiments, such as NO$\nu$A \cite{NOvA0503,NOvA1209}, see \cite{hierarchy} for more details. For the atmospheric angle, PINGU and MINOS \cite{minos1206,minos1207} can help to narrow its uncertainty and hence tell its deviation from $45^\circ$ and to which side, or octant, it will deviate. The {\tt CP} effect can be determined by accelerator type experiments such as T2K \cite{T2K0106} and NO$\nu$A. These experiments will report data in the next few years and make precision tests of neutrino mixing models possible.

In this paper, we explore the phenomenological implications of the residual $\mathbb Z^s_2$ and $\overline{\mathbb Z}^s_2$ symmetries in detail. In \gsec{sec:prediction}, the unique correlation between the neutrino mixing parameters is used to predict the Dirac {\tt CP} phase and the atmospheric angle with one of the recent global fits. In \gsec{sec:simulation} these predictions are then used as input to future precision neutrino experiments to study the prospects of verifying/falsifying the $\mathbb Z^s_2$ and $\overline {\mathbb Z}^s_2$ residual symmetries. We will conclude the paper in \gsec{sec:conclusion}.

\section{Predictions of $\mathbb Z^s_2$ and $\overline {\mathbb Z}^s_2$}
\label{sec:prediction}

After the $\mu$--$\tau$ symmetry is broken, the remaining residual symmetry of the neutrino sector can either be $\mathbb Z^s_2$ or $\overline{\mathbb Z}^s_2$, if neutrinos are of the Majorana type. Each of them can induce a unique correlation among the mixing parameters, namely the three mixing angles $\theta_r (\equiv \theta_{13})$, $\theta_s (\equiv \theta_{12})$, and $\theta_a (\equiv \theta_{23})$ together with the Dirac {\tt CP} phase $\delta_D$, as follows \cite{Ge1104, Ge1108}, 
\begin{subequations}
\begin{eqnarray}
  \cos \delta_D 
& = &
  \frac{(s_s^2 - c_s^2 s_r^2)(s_a^2 - c_a^2)}{4 c_a s_a c_s s_s s_r}
\qquad 
  \mbox{for} \quad \mathbb Z^s_2 \,,
\label{eq:cD_a}
\\
  \cos \delta_D 
& = &
  \frac{(s_s^2 s_r^2 - c_s^2)(s_a^2 - c_a^2)}{4 c_a s_a c_s s_s s_r}
\qquad 
  \mbox{for} \quad \overline{\mathbb Z}^s_2 \,,
\label{eq:cD_b}
\end{eqnarray}
\label{eq:cD}
\end{subequations}
\hspace{-2.5mm}
where $(c_\alpha, s_\alpha) \equiv (\cos \theta_\alpha, \sin \theta_\alpha)$ and the subscripts are chosen according to the physical/historical meaning of the corresponding mixing angles. Note that $\theta_r$ is denoted as $\theta_x$ in \cite{Ge1104,Ge1108} because at that time it had not yet been measured, but now history has marked it as the one first measured by reactor neutrino experiments. In addition, the PDG convention of the mixing matrix \cite{pdg} has been adopted resulting in a minus sign for the expressions of $\cos \delta_D$. The above correlations and the expanded forms \cite{Ge1104,*Ge1108} are reproduced in various models, \cite{cited1,*cited2,*cited3,*cited4,Smirnov1204,Smirnov1212} and \cite{old1,*expand1,*expand2,*expand3,*expand4,*expand5}, respectively.

There is an important property of the above correlations. These expressions contain only mixing parameters with no reference to model parameters. In other words, it is a unique prediction that can serve as a robust indication of the existence of residual $\mathbb Z^s_2$ or $\overline{\mathbb Z}^s_2$ symmetries and can be directly verified by precision measurements. This property make the correlation very restrictive and powerful. 

With the reactor angle $\theta_r$ around $8.7^\circ$ \cite{Daya1210}, the Dirac {\tt CP} phase (\ref{eq:cD}) has a large chance to fall into the meaningful range of $\cos \delta_D \in [-1,1]$. If the atmospheric angle $\theta_a$ also deviates from its maximal value $\theta_a = 45^\circ$, as indicated by global fits \cite{Fogli1205,Maltoni1209} and the preliminary measurements of MINOS \cite{minos1206,*minos1207}, a reasonable prediction of $\delta_D$ can be obtained \cite{Ge1104,*Ge1108}. Since our last paper, the global fits have been updated by including the recent measurements from reactor neutrino experiments \cite{Valle1205,Fogli1205,Maltoni1209} and it would be interesting to see to what extent the new data affect the prediction. 

In order to make a close comparison with the results shown in \cite{Ge1108}, we adopt the global fit updated in \cite{Fogli1205}. The $\chi$ functions concerning the six neutrino oscillation parameters have been summarized in Fig.~3 therein. Of these six parameters, the solar mass squared difference $\delta m^2_s (\equiv \delta m^2_{12})$, the solar angle $\theta_s$, and the reactor angle $\theta_r$ have $\chi$ functions that are independent of the mass hierarchy, while the constraint on the absolute value of the atmospheric mass squared difference $\delta m^2_a (\equiv \delta m^2_{13})$ has a slight dependence. The largest difference comes from the atmospheric angle $\theta_a$ and the Dirac {\tt CP} phase $\delta_D$ due to the octant-hierarchy and CP-hierarchy degeneracies \cite{degeneracy01,degeneracy04}. The resolution of the atmospheric angle's octant is better for normal hierarchy (NH) than for inverted hierarchy (IH). Since the correlations (\ref{eq:cD}) are functions of $\theta_a$ and $\delta_D$, the resulting predictions will also bear some dependence on the neutrino mass hierarchy. To make it more realistic, we extract \footnote{With the scientific data extraction tool {\it g3data}, http://www.frantz.fi/software/g3data.php.} the exact $\chi$ curves from Fig.3 of \cite{Fogli1205} as input. In this way, the complicated distributions, especially that of $\theta_a$, and the mass hierarchy dependence are both taken into account in contrast to using the (a)symmetric Gaussian distribution in \cite{Ge1104,*Ge1108}.

By combining the correlations (\ref{eq:cD}) and the global fit \cite{Fogli1205}, we can obtain a distribution for $\cos \delta_D$ using the following integration,
\begin{equation} 
  \frac{d P(\cos \delta_D)}{d \cos \delta_D} 
=
  \int \delta^p_D 
  \mathbb P(s_a^2) 
	\mathbb P(s_s^2) 
	\mathbb P(s_r^2) 
	\mathrm{d}s_a^2 \mathrm{d}s_s^2 \mathrm{d}s_r^2 \,,
\label{eq:originalDist}
\end{equation}
where $\delta^p_D \equiv \delta(\cos{\delta_D} - \overline{c}_D)$ is a $\delta$-function with $\overline{c}_D$ denoting the RHS of (\ref{eq:cD}), and the $\mathbb{P}$ function denotes the normalized distribution which is related to the $\chi$ function of the corresponding parameter in \cite{Fogli1205} as $P \propto \exp(-\chi^2/2)$. The integration (\ref{eq:originalDist}) is carried out by an adapted C++ version of the Monte Carlo integration and event generation package BASES \cite{bases86,*bases95,*bases11}. Then, the distribution of the Dirac {\tt CP} phase can be obtained through,
\begin{equation} 
  \frac{d P(\delta_D)}{d \delta_D} 
=
  |s_D| \frac{d P(\cos \delta_D)}{d \cos \delta_D} \,.
\label{eq:changeVar}
\end{equation}
In principle, the $\chi^2(\delta_D)$ function for the Dirac {\tt CP} phase in Fig.~3 of \cite{Fogli1205} can also be extracted and implemented to account for the prior constraint. In that case, there would be an extra $\mathbb P(\delta_D)$ in (\ref{eq:originalDist}). Nevertheless, it has negligible effect on the prediction, and is neglected in the following discussions with the only exception being in the prediction of the atmospheric angle.

\begin{figure}[h!]
\centering
\includegraphics[width=0.5\textwidth,height=5.5cm]{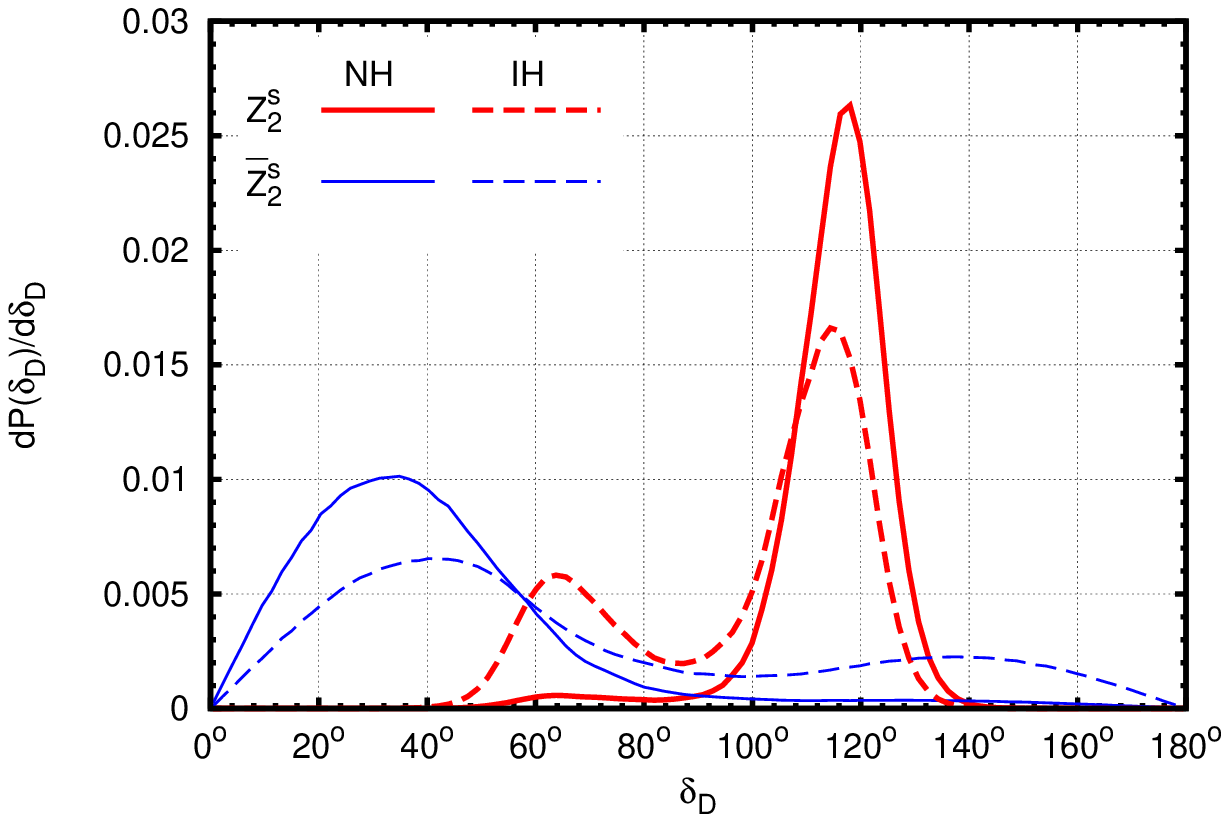}
\includegraphics[width=0.5\textwidth,height=5.5cm]{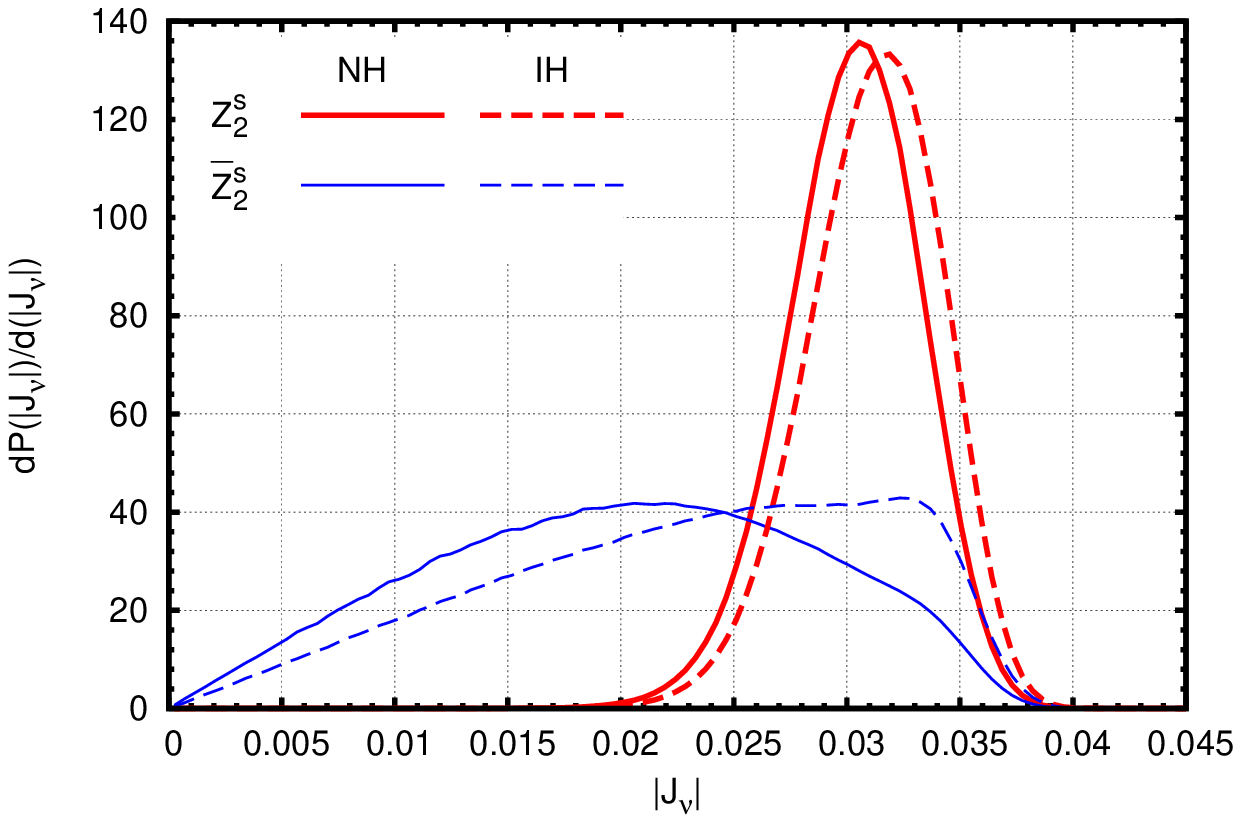}
\caption{Predicted distributions of the Dirac {\tt CP} phase $\delta_D$ and the leptonic Jarlskog invariant $J_\nu$.}
\label{fig:dCP}
\end{figure}

The predicted differential probability distributions of the Dirac {\tt CP} phase $\delta_D$ and the Jarlskog invariant $J_\nu \equiv c_a s_a c_s s_s c^2_r s_r \sin \delta_D$ \cite{Jarlskog85,*Jarlskog11} are shown in the upper and lower panels of \gfig{fig:dCP}, respectively. Note that for a certain value of $\cos \delta_D$, there are two solutions of which only the one in the range $0 < \delta_D < \pi$ is displayed. There is a mirror distribution $\delta_D \rightarrow - \delta_D$ that should be kept in mind. So the curves in the upper panel of \gfig{fig:dCP} have an extra normalization factor of $0.5$ in addition to the expression in (\ref{eq:originalDist}). For the Jarlskog invariant $J_\nu$, the two mirror distributions have been combined by displaying $|J_\nu|$ instead of $J_\nu$.

For the distribution of $\delta_D$ shown in \gfig{fig:dCP}, there is only one prominent peak for NH, whereas an extra peak appears for IH due to the strong dependence of the input $\chi^2(s^2_a)$ function on the mass hierarchy. Note that the two peaks have approximately equal distances to the middle point $\delta_D = 90^\circ$, since the two local minima in $\chi^2(s^2_a)$ sit at the two sides of $s^2_a = 0.5$ symmetrically. If only the prominent peaks are considered, the predicted $\delta_D$ distribution of $\mathbb Z^s_2$ can be clearly distinguished from that of $\overline{\mathbb Z}^s_2$. In other words, if the neutrino mass hierarchy is normal these two residual symmetries can be discriminated by simply measuring the Dirac {\tt CP} phase precisely. The dependence on other parameters is elaborated in \gsec{sec:simulation}. For IH, the minor peak of $\mathbb Z^s_2$ has some overlap with the major peak of $\overline{\mathbb Z}^s_2$. Thus, there is still a small chance that measuring only the Dirac {\tt CP} phase will not suffice to differentiate between the two symmetries. 

In spite of the apparent difference between the $\delta_D$ distributions of $\mathbb Z^s_2$ with NH and IH, the distributions of the Jarlskog invariant are quite close to each other. This is because the minor peak for IH is actually a mirror to the major one with $\delta_D \rightarrow 180^\circ - \delta_D$, and the major peaks for NH and IH overlap with one another. It is difficult to distinguish the mass hierarchy if only the leptonic Jarlskog invariant $J_\nu$ can be measured. This is also true for $\overline{\mathbb Z}^s_2$ as the two curves for NH and IH are not so far from each other in most regions. A vacuum oscillation experiment may not be able to measure the difference, as the {\tt CP} effect is always proportional to the leptonic Jarlskog invariant. However, the difference between $\mathbb Z^s_2$ and $\overline{\mathbb Z}^s_2$ is significant. The peaks of the former are very narrow, while those of the latter extend through the whole range from $0$ to $0.04$. This is different from the results in \cite{Ge1108} using the global fit before the reactor angle had been measured. It shows that precision measurements of the mixing parameters can really help to distinguish residual symmetries. If future measurements tell us that $0.02 < |J_\nu| < 0.04$, $\mathbb Z^s_2$ would become the most likely residual symmetry, while for $|J_\nu| < 0.02$ only $\overline{\mathbb Z}^s_2$ can survive. If the true value of $|J_\nu|$ is even larger than $0.04$, both of them can be eliminated.

\begin{figure}[t!]
\centering
\includegraphics[width=0.5\textwidth,height=5.5cm]{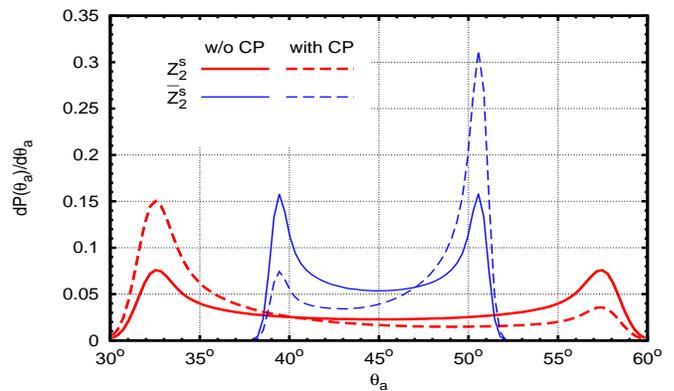}
\caption{Predicted distributions of the atmospheric angle.}
\label{fig:ta}
\end{figure}

With the reactor angle being precisely measured, the atmospheric angle becomes the parameter in need of improvement, especially its deviation from $45^\circ$. The correlation (\ref{eq:cD}) can now be used to predict the atmospheric angle $\theta_a$ as a function of the reactor angle $\theta_r$, the solar angle $\theta_s$, and the Dirac {\tt CP} phase $\delta_D$, in a similar way as (\ref{eq:originalDist}) and (\ref{eq:changeVar}). The results are shown in \gfig{fig:ta}. Since the global fit results of $\theta_r$, $\theta_s$, and $\delta_D$ have little dependence on the neutrino mass hierarchy, the predicted distribution of $\theta_a$ is almost the same between NH and IH. However, the result depends on the distribution of $\delta_D$. Without any constraint on $\delta_D$, the predicted $\theta_a$ sits symmetrically on the two sides of $\theta_a = 45^\circ$, peaking around $32^\circ \sim 33^\circ$ or $57^\circ \sim 58^\circ$ for $\mathbb Z^s_2$, and $39^\circ$ or $51^\circ$ for $\overline{\mathbb Z}^s_2$. When imposing a prior on $\delta_D$, such as the global fit \cite{Fogli1205} where $\delta_D \approx 180^\circ$ is favored, the predicted distributions of $\theta_a$ become asymmetric around the middle point. This is because a prior favoring $\delta_D \approx 180^\circ$ brings a preferred sign into $\tan 2 \theta_a$ through $\cos \delta_D$ in (\ref{eq:cD}). This sign combines with the mainly positive factor $(s^2_s - c^2_s s^2_r)$ (\ref{eq:cD_a}) for $\mathbb Z^s_2$ and the mainly negative factor $(s^2_s s^2_r - c^2_s)$ (\ref{eq:cD_b}) for $\overline{\mathbb Z}^s_2$ to account for the major peak in the lower octant (LO) or the higher octant (HO), respectively. Things can be different if other priors on $\delta_D$, such as \cite{Valle1205} and \cite{Maltoni1209}, are imposed. It is essential to have a precision measurement on $\delta_D$. In addition, the distribution is constrained within $30^\circ < \theta_a < 60^\circ$ for $\mathbb Z^s_2$ and $38^\circ < \theta_a < 52^\circ$ for $\overline{\mathbb Z}^s_2$. If a large enough deviation of $\theta_a$ is observed, $\overline{\mathbb Z}^s_2$ can be immediately excluded.

With a precisely measured reactor angle, the predicted distributions of the Dirac {\tt CP} phase and the atmospheric angle are quite different for $\mathbb Z^s_2$ and $\overline{\mathbb Z}^s_2$. This shows the possibility of distinguishing between them at neutrino experiments, as described in \gsec{sec:simulation}

\section{Testing $\mathbb Z^s_2$ and $\overline{\mathbb Z}^s_2$ with Precision Experiments}
\label{sec:simulation}

Since the correlations (\ref{eq:cD}) involve the three mixing angles and the Dirac {\tt CP} phase, it is necessary for all of them to be precisely measured in order to finally verify the residual symmetry. According to \gsec{sec:prediction}, the current $\chi^2(\delta_D)$ function obtained indirectly from the global fit does not have much effect, especially in comparison with the prediction of $\delta_D$ itself from (\ref{eq:cD}). A more precise measurement of $\delta_D$ is necessary.

In the near future, the Dirac {\tt CP} phase will be measured by the two experiments T2K \cite{T2K0106} and NO$\nu$A \cite{NOvA0503,NOvA1209}. So we will focus on these two experiments to explore the phenomenological consequences of the residual $\mathbb Z^s_2$ and $\overline{\mathbb Z}^s_2$ symmetries. The predictions (\ref{eq:cD}) of the Dirac {\tt CP} phase in \gsec{sec:prediction}, together with the normalized global fit distributions of the remaining five parameters \cite{Fogli1205}, are used as input for simulation with GLoBES \cite{globes04,*globes07} using the model files for T2K \cite{t2kglb1,*t2kglb2,*cx} and NO$\nu$A \cite{NOvA0503,novasignal} with cross sections taken from \cite{flux1,*flux2}. There are three major modifications: 
\begin{enumerate}
\item The matter density distribution \cite{t2kdensity} is adopted for T2K to take the complicated geological structure of Japan into consideration. For NO$\nu$A, we use the default constant matter density $\rho = 2.8 g/cm^2$. In addition, each experiment has a 5\% uncertainty in the normalization of the matter density.

\item As pointed out in \cite{equalrunning}, splitting the running time equally among neutrinos and antineutrinos would help to avoid the chance of failing to identify the true hierarchy at the NO$\nu$A experiment. So we choose to split the $3$-year run of NO$\nu$A and the $5$-year run of T2K equally among neutrinos and antineutrinos. As a comparison, the schemes of purely neutrinos or antineutrinos are also explored.

\item User-defined priors are implemented instead of the default Gaussian distribution. To comply with this flexibility, the minimization and projection of $\chi^2$ functions are carried out by the external package, MINUIT2 \cite{minuit2}.
\end{enumerate}

Since the Dirac {\tt CP} phase is a function of the mixing angles now, only two major degrees of freedom, namely the neutrino mass hierarchy and the octant of the atmospheric angle, need to be explored at neutrino experiments. The other four parameters, the two mass squared differences, the solar angle, and the reactor angle, are set to be their corresponding best fit values \cite{Fogli1205},
\begin{subequations}
\begin{eqnarray}
  \delta m^2 
=
  7.54 \times 10^{-5} \mbox{eV}^2
& \quad &
  \mbox{(NH \& IH)} \,,
\\
  \Delta m^2
=
  2.43 \mbox{ or } 2.42 \times 10^{-3} \mbox{eV}^2 
& \quad &
  \mbox{(NH or IH)} \,,
\\
  \sin^2 \theta_s 
=
  0.307 
& \quad &
  \mbox{(NH \& IH)} \,,
\\
  \sin^2 \theta_r
=
  0.0241 \mbox{ or } 0.0244
& \quad &
  \mbox{(NH or IH)} \,.
\quad
\end{eqnarray}
\label{eq:inputs}
\end{subequations}
\hspace{-2.2mm}
Note that the mass difference in the $\chi^2(|\Delta m^2|)$ function is defined as $\Delta m^2 \equiv m^2_3 - (m^2_1 + m^2_2)/2$ \cite{Fogli1205} which needs to be converted into $\delta m^2_{31} \equiv m^2_3 - m^2_1$ before being put into GLoBES. The neutrino mass hierarchy, $m^2_3 > m^2_1$ (NH) or $m^2_3 < m^2_1$ (IH), as well as the Dirac {\tt CP} phase $\delta_D$ and the atmospheric angle $\theta_a$ are tuned to generate pseudo data which is fit by minimizing the $\chi^2$ function,
\begin{equation}
  \chi^2
\equiv
  \chi^2_{\rm stat} 
+ \chi^2(\delta m^2)
+ \chi^2(|\Delta m^2|)
+ \chi^2(s^2_s)
+ \chi^2(s^2_r) \,.
\label{eq:chi2}
\end{equation}
The first term represents the statistical contribution from the event rates registered by the experiments under consideration, and the following four terms are the priors extracted from the global fit \cite{Fogli1205}. Note that the priors on the atmospheric angle $\chi^2(s^2_a)$ and the Dirac {\tt CP} phase $\chi^2(\delta_D)$ are not included, in order to show the pure sensitivity from new measurements without contamination of their priors.

\subsection{The Neutrino Mass Hierarchy}

Since the neutrino mass hierarchy is a discrete degree of freedom, we can use either NH or IH to fit the pseudo data and obtain the corresponding minimum $\chi^2_{min}$. If the mass hierarchy used to fit the pseudo data is the same as the one used to generate it, the $\chi^2$ function (\ref{eq:chi2}) can be minimized to zero. Otherwise, a nonzero minimum would result. The difference represents the sensitivity of distinguishing NH from IH. For convenience, we show its absolute value, $\Delta \chi^2 \equiv |\chi^2_{min}(NH) - \chi^2_{min}(IH)|$, where NH and IH stand for the mass hierarchy used to fit the data. 

\begin{figure}[h!]
\includegraphics[height=0.5\textwidth,width=5.5cm,angle=-90]{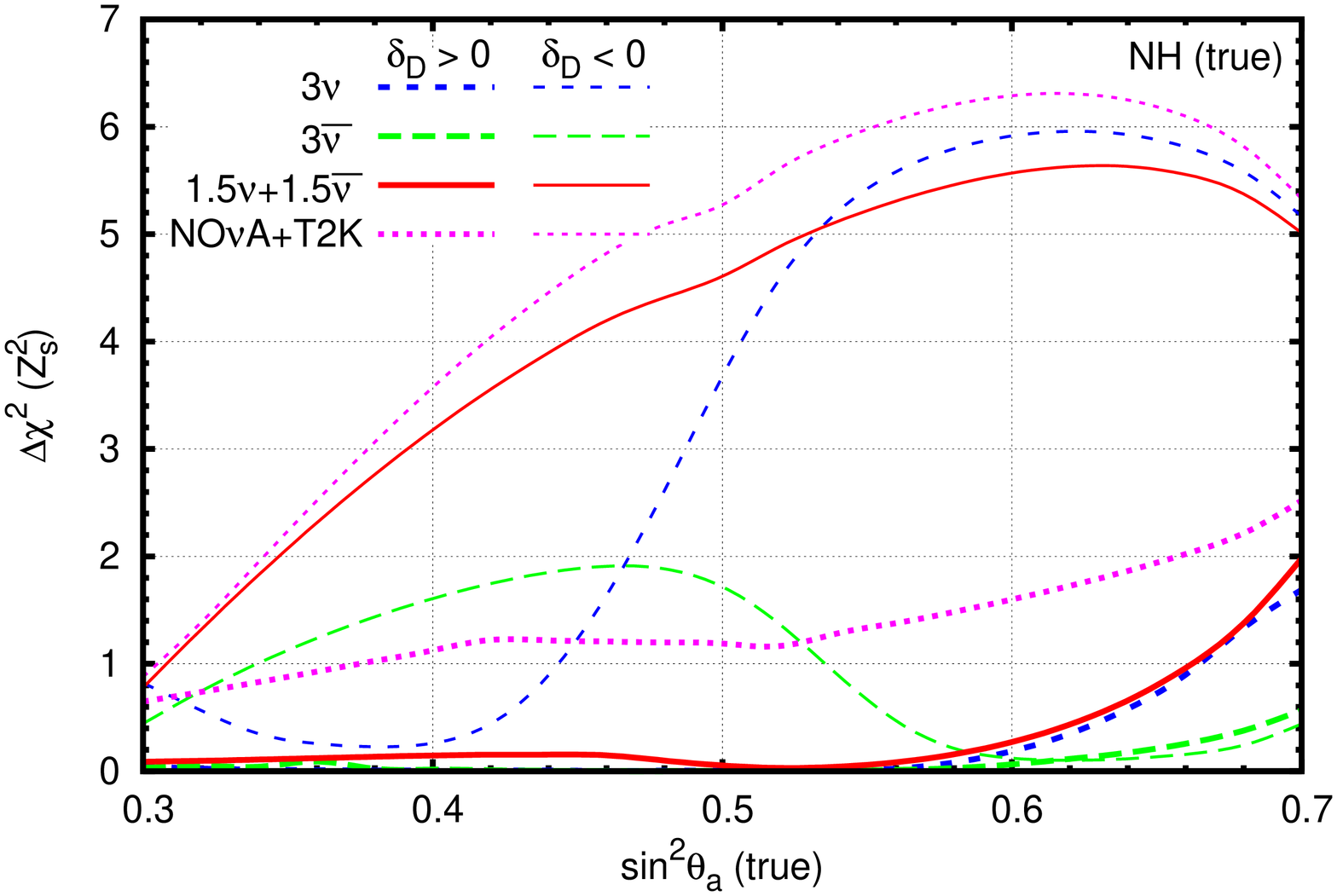}
\includegraphics[height=0.5\textwidth,width=5.5cm,angle=-90]{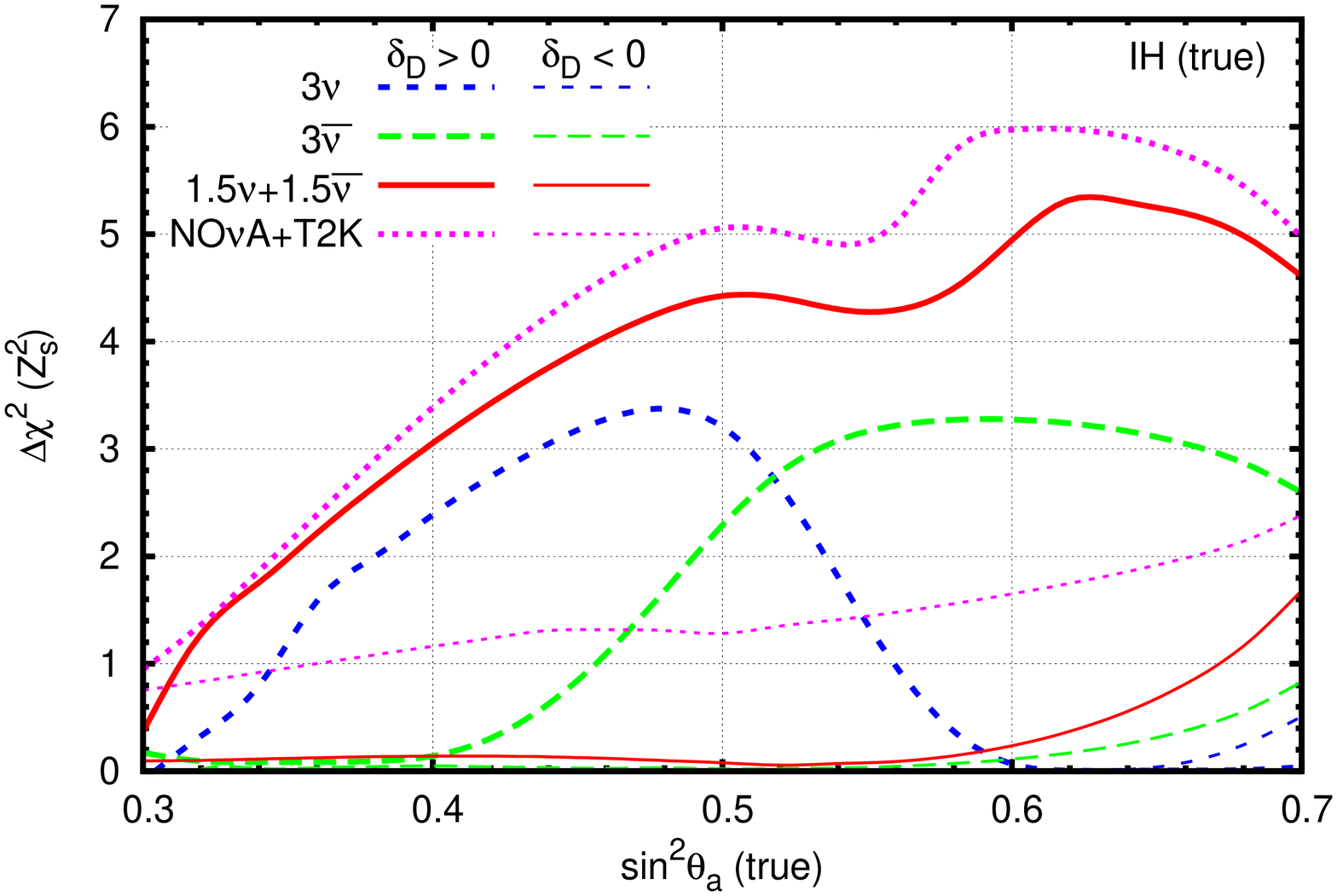}
\caption{Hierarchy sensitivity for $\mathbb Z^s_2$.}
\label{fig:MH-Z2}
\end{figure}

\begin{figure}[h!]
\centering
\includegraphics[height=0.5\textwidth,width=5.5cm,angle=-90]{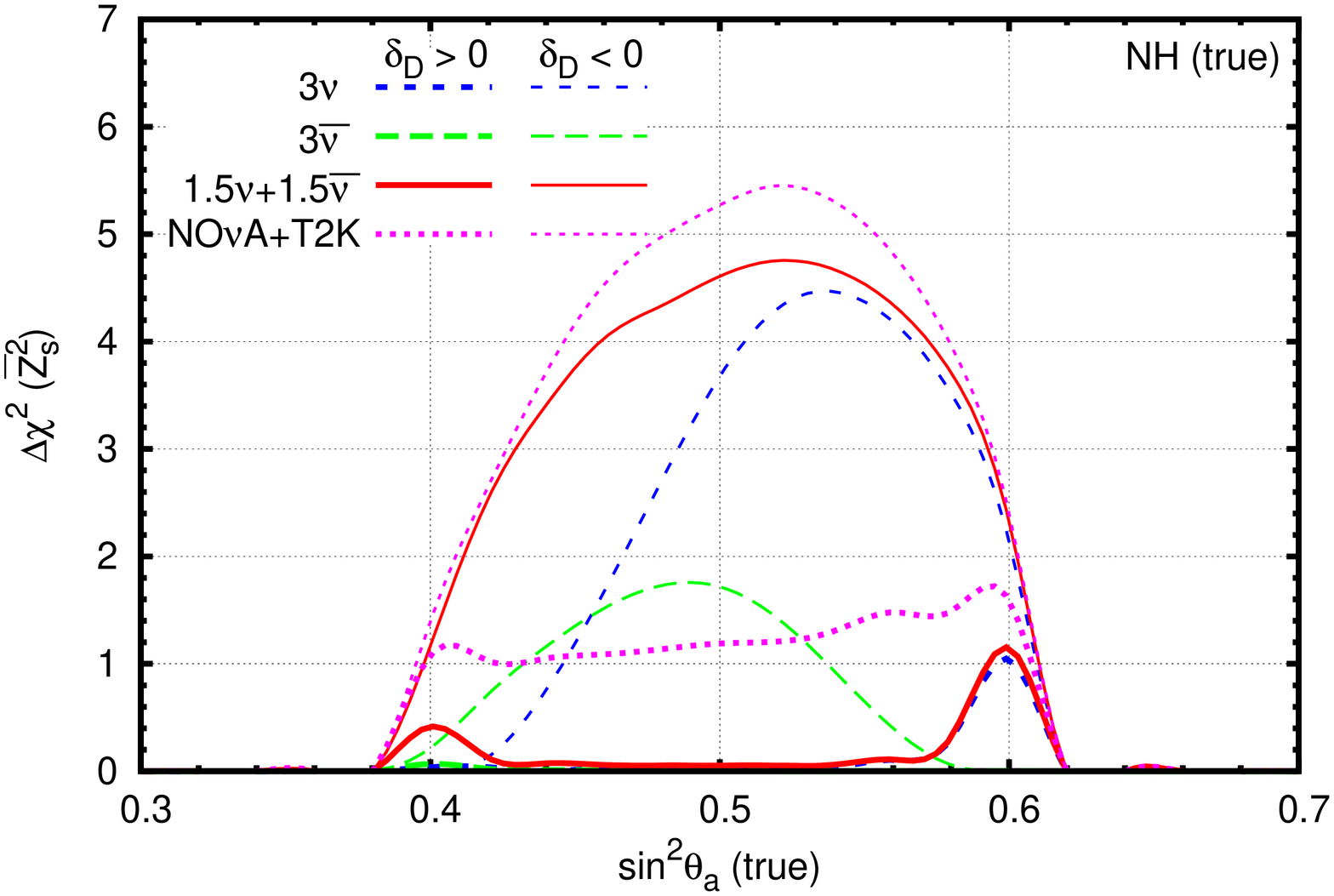}
\includegraphics[height=0.5\textwidth,width=5.5cm,angle=-90]{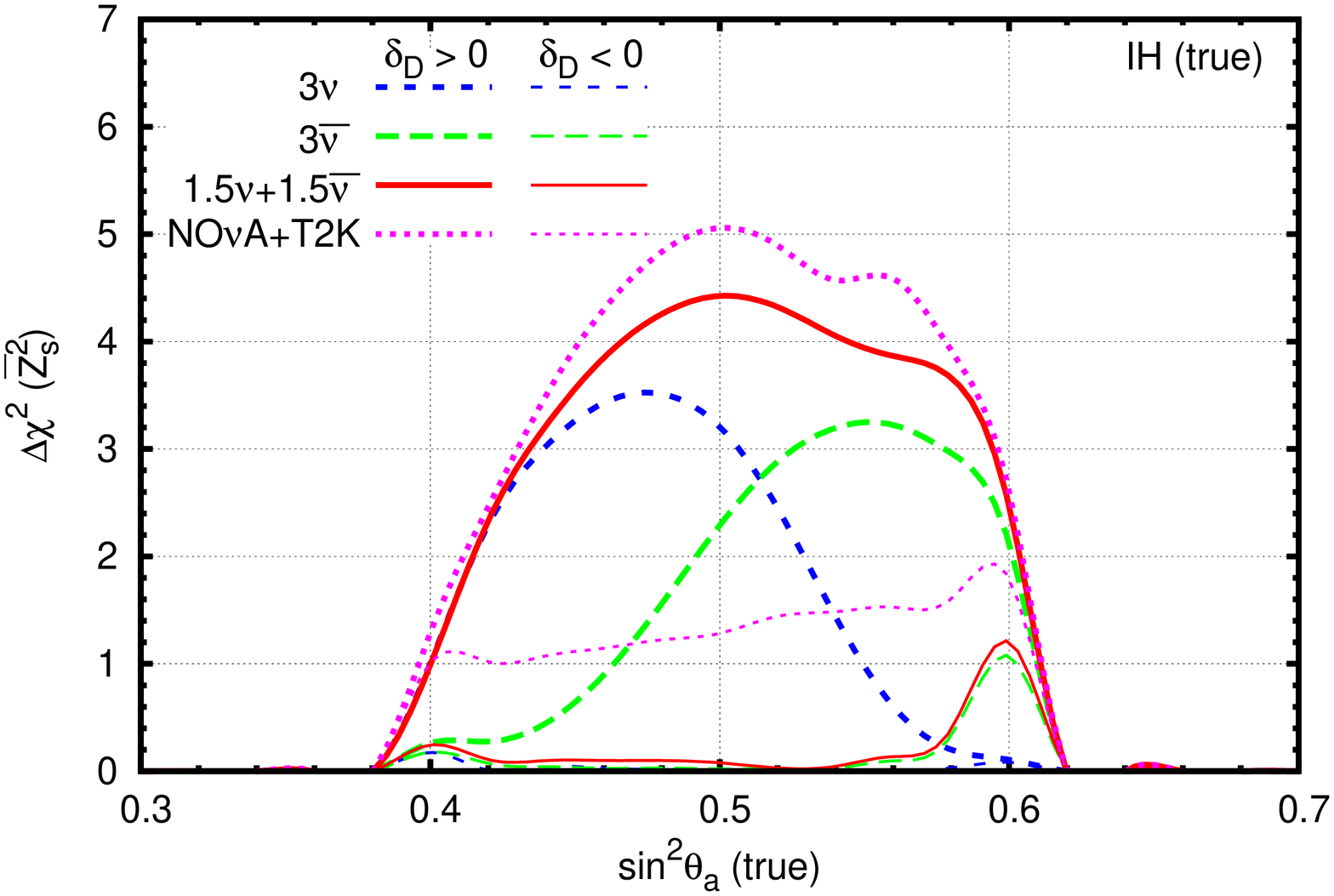}
\caption{Hierarchy sensitivity for $\overline{\mathbb Z}^s_2$.}
\label{fig:MH-Z2bar}
\end{figure}

In addition to the input values in (\ref{eq:inputs}), the Dirac {\tt CP} phase $\delta_D$ is expressed as a function of $\theta_a$ (\ref{eq:cD}) with two degenerate solutions $\delta_D \in [0^\circ, 180^\circ]$ and $\delta_D \in [-180^\circ, 0^\circ]$. These are used as true values to generate the pseudo data to which the $\chi^2$ fit is carried out with the six neutrino oscillation parameters being free. The results are shown in \gfig{fig:MH-Z2} and \gfig{fig:MH-Z2bar} as functions of the true value of $\theta_a$. The negative ($\delta_D < 0$) and positive ($\delta_D > 0$) solutions are plotted in thin and thick curves. To see the benefit of splitting the running time among neutrinos and antineutrinos, three schemes, namely NO$\nu$A with 3-years of neutrinos ($3\nu$), or 3-years of antineutrinos ($3 \bar \nu$), or 1.5-years of neutrinos and antineutrinos ($1.5 \nu + 1.5 \bar \nu$), have been explored. Finally, we show the combined result (NO$\nu$A+T2K), with 1.5 years at NO$\nu$A and $2.5$ years at T2K, each for neutrinos and antineutrinos.

\gfig{fig:MH-Z2} shows the hierarchy sensitivities when $\mathbb Z^s_2$ is imposed. For NH with $\delta_D > 0$ and IH with $\delta_D < 0$, there is almost no chance to identify the neutrino mass hierarchy, due to CP-hierarchy degeneracies \cite{degeneracy01,*degeneracy04}. The following discussions focus on the other two cases, NH with $\delta_D < 0$ and IH with $\delta_D > 0$. If NO$\nu$A runs for 3 years with neutrinos ($3 \nu$) or antineutrinos ($3 \bar \nu$), the mass hierarchy can be identified in only a part of the region of $\sin^2 \theta_a$. For NH with $\delta_D < 0$, the sensitivity $\Delta \chi^2 > 4$ can be reached for $0.5 < \sin^2 \theta_a < 0.7$ with a 3-year run of neutrinos, while it can never reach $\Delta \chi^2 > 2$ in the entire region if a 3-year run of antineutrinos is adopted. The situation can be significantly improved if NO$\nu$A runs with an equal splitting of the running time between the neutrinos and antineutrinos ($1.5 \nu + 1.5 \bar \nu$). The sensitivity can increase to $\Delta \chi^2 > 2$ for almost all values of $\sin^2 \theta_a$, and the coverage of $\Delta \chi^2 > 4$ now extends to $\sin^2 \theta_a > 0.45$ and even $\sin^2 \theta_a > 0.41$ if combined with T2K (NO$\nu$A+T2K). A similar thing happens for IH with $\delta_D > 0$. The only difference is, if running with only neutrinos or antineutrinos, the sensitive octant switches in comparison with the case of NH with $\delta_D < 0$. The sensitivity reaches $\Delta \chi^2 > 2$ in the region $0.37 < \sin^2 \theta_a < 0.54$ with a neutrino run and $\sin^2 \theta_a > 0.49$ with an antineutrino run.  In either case, splitting the running time helps to avoid the uncertainty from the unknown octant of $\theta_a$. Note that, T2K can contribute at most $\Delta \chi^2 \sim 1$ to the measurement of the neutrino mass hierarchy. 

The results for $\overline{\mathbb Z}^s_2$ are shown in \gfig{fig:MH-Z2bar}. Most features of $\mathbb Z^s_2$ still apply here. Nevertheless, the differences are also apparent. First, not every value of $\sin^2 \theta_a$ corresponds to a Dirac {\tt CP} phase $\delta_D$ through (\ref{eq:cD}). For $\sin^2 \theta_a < 0.38$ and $\sin^2 \theta_a > 0.62$, the prediction runs out of the meaningful range, $-1 < \cos \delta_D < 1$. Actually, its inverse has already been observed in \gfig{fig:ta} and the discussions there. When $\theta_a$ approaches the endpoints, $\cos \delta_D$ approaches the crossing point, $\cos \delta_D = \pm 1$, between the upper half plane, $0^\circ < \delta_D < 180^\circ$, and the lower half plane, $-180^\circ < \delta_D < 0^\circ$. The pair of curves with $\delta_D > 0$ and $\delta_D < 0$ would converge there. Due to CP-hierarchy degeneracies \cite{degeneracy01,*degeneracy04}, the sensitivity $\Delta \chi^2$ approaches zero when they converge. Another difference from \gfig{fig:MH-Z2} is that the sensitivity of distinguishing NH and IH is slightly smaller for $\overline{\mathbb Z}^s_2$.

\subsection{Distinguishing $\mathbb Z^s_2$ and $\overline{\mathbb Z}^s_2$}
\label{sec:cp}

Once the neutrino mass hierarchy is determined, the next question is how to distinguish between $\mathbb Z^2_s$ and $\overline{\mathbb Z}^s_2$. For this purpose, the six parameters used to do the $\chi^2$ fit are no longer independent of each other. One degree of freedom can be removed by the correlation (\ref{eq:cD}). The corresponding sensitivity $\Delta \chi^2 \equiv |\chi^2_{min} (\mathbb Z^s_2) - \chi^2_{min} (\overline{\mathbb Z}^s_2)|$ is defined as a function of $\delta_D$, in the same way as the sensitivity to the mass hierarchy. 

\begin{figure}[h!]
\centering
\includegraphics[width=5.5cm,height=0.5\textwidth,angle=-90]{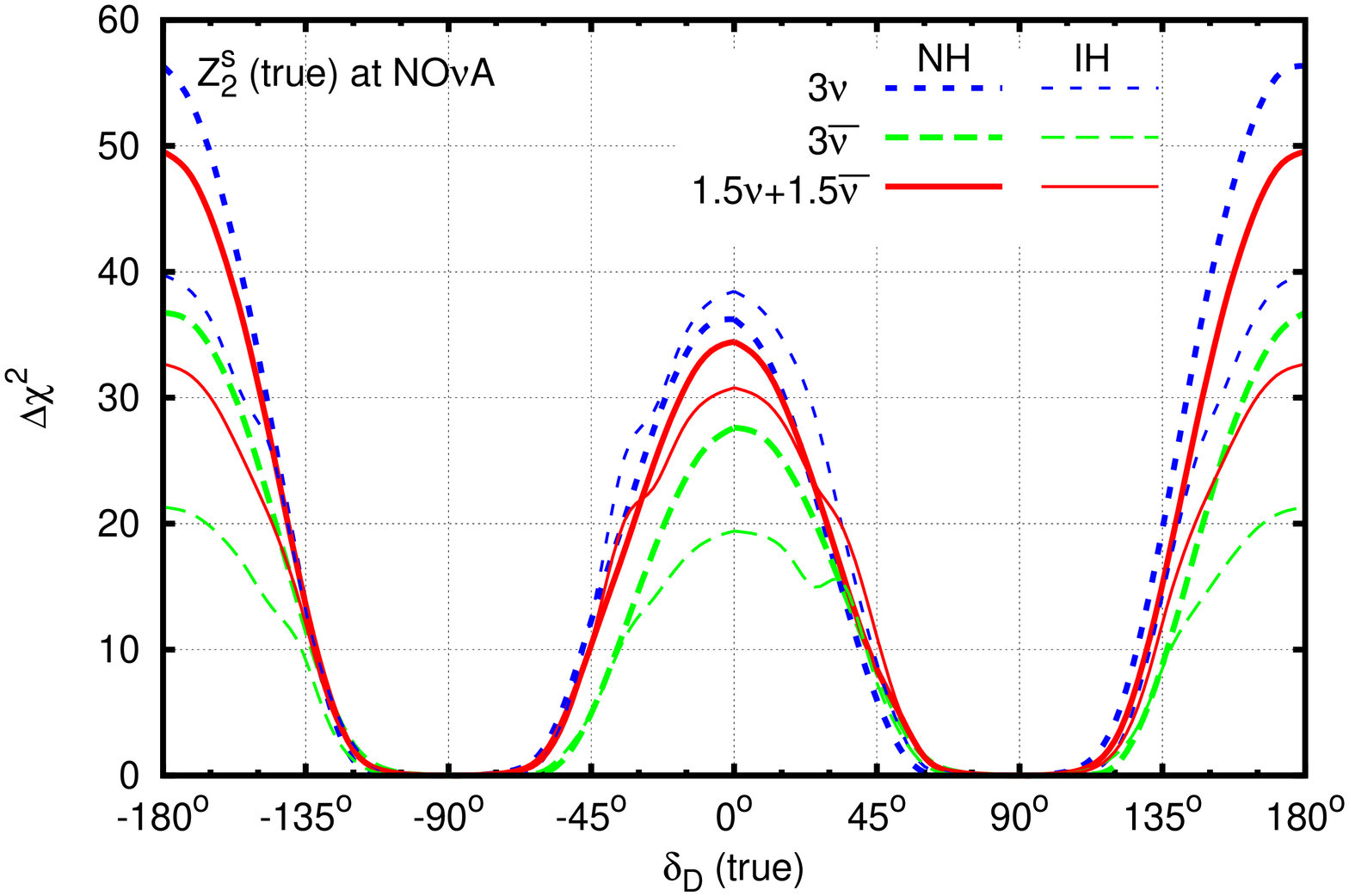}
\includegraphics[width=5.5cm,height=0.5\textwidth,angle=-90]{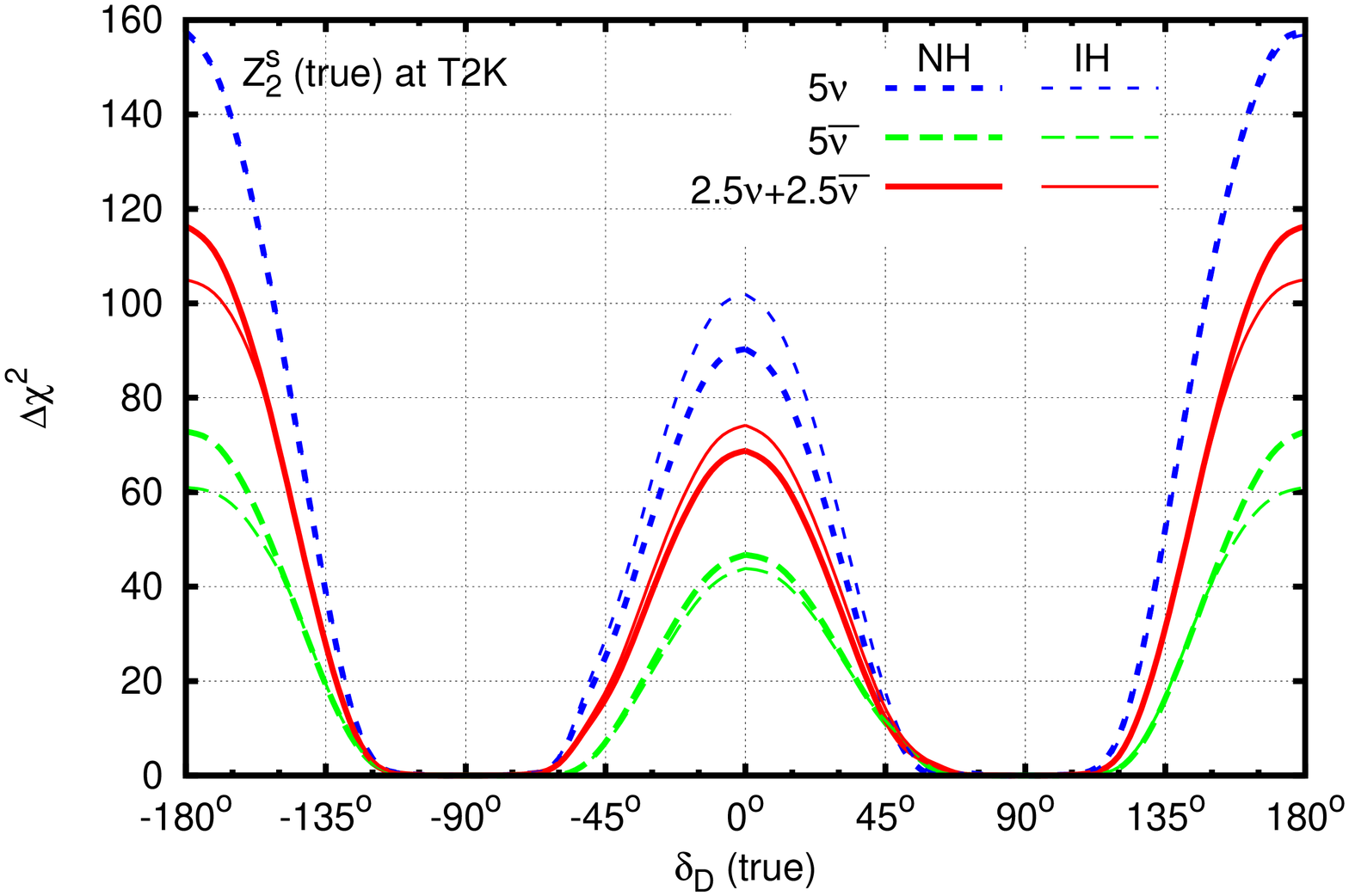}
\caption{Sensitivity of distinguishing $\mathbb Z^s_2$ and $\overline{\mathbb Z}^s_2$.}
\label{fig:z2s}
\end{figure}

In \gfig{fig:z2s}, we show the results obtained by generating the pseudo data with $\mathbb Z^s_2$ and fitting it with $\overline{\mathbb Z}^s_2$. The first thing to be noticed is the oscillatory behavior. For $\delta_D \approx \pm 90^\circ$, no difference between the two residual symmetries can be observed. This is because, around these two places, $\cos \delta_D$ is very close to zero, leading to an almost maximal atmospheric angle for both $\mathbb Z^s_2$ and $\overline{\mathbb Z}^s_2$. It can not be changed by adjusting other parameters. For $\cos \delta_D \approx \pm 1$, running with neutrinos is always better than running with antineutrinos. As a consequence, splitting the running time would compromise some sensitivity, but it is still acceptable, since the sensitivity is already large enough. In addition, the measurement is easier if the true mass hierarchy is inverted for $\delta_D \approx 0^\circ$ and normal for $\delta_D \approx 180^\circ$. The above observation also applies to T2K, as shown in \gfig{fig:z2s} with the only difference that now NH is the one that has larger sensitivity for a 5-year run of neutrinos.

If $\overline{\mathbb Z}^s_2$ is the true residual symmetry, fitting with $\mathbb Z^s_2$ would not achieve any remarkable sensitivity at NO$\nu$A and T2K. The reason for this appears in \gfig{fig:ta}. The distribution of $\theta_a$ implied by $\mathbb Z^2_2$ can cover the whole distribution permitted by $\overline{\mathbb Z}^s_2$. So it is much easier for $\mathbb Z^s_2$ to fit $\overline{\mathbb Z}^s_2$.

\subsection{The Octant of $\theta_a$}

As shown in \gfig{fig:z2s}, the octant of the atmospheric angle is also essential to distinguish $\mathbb Z^s_2$ and $\overline{\mathbb Z}^s_2$. From (\ref{eq:cD}) we can see that, if the atmospheric angle $\theta_a$ is not maximal, the Dirac {\tt CP} phase will deviate from $\delta_D = \pm 90^\circ$ and the sensitivity of distinguishing $\mathbb Z^s_2$ and $\overline{\mathbb Z}^s_2$ can increase significantly. 

\begin{figure}[t]
\hfill
\includegraphics[height=0.45\textwidth,width=8cm,angle=-90]{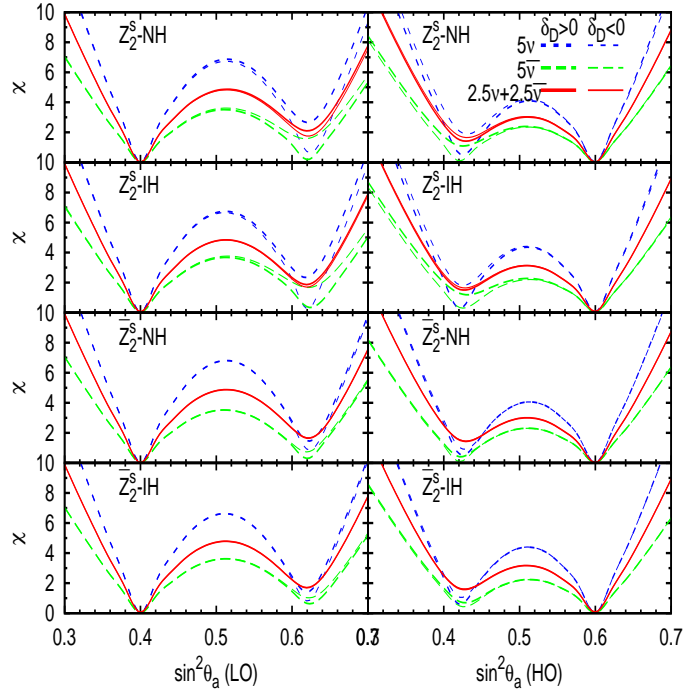}
\vspace{1cm}
\caption{Octant Sensitivity for $\mathbb Z^s_2$ and $\overline{\mathbb Z}^s_2$ at T2K.}
\label{fig:octant}
\end{figure}

In \gfig{fig:octant} we show the ability of T2K to measure the octant of the atmospheric angle. In the simulation, the correlation (\ref{eq:cD}) of $\mathbb Z^s_2$ or $\overline{\mathbb Z}^s_2$ is implemented to generate the pseudo data with $s^2_a = 0.4$ for LO and $s^2_a = 0.6$ for HO. The $\chi^2$ fit is carried out with the atmospheric angle $\theta_a$ fixed while the other 5 parameters can be freely adjusted. 

It is a general feature that there are two local minima, one at the input value of $s^2_a$ and the other at its mirror $s^2_a \rightarrow 1 - s^2_a$. The octant sensitivity can be parametrized by the difference between these two local minima $\Delta \chi^2 \equiv |\chi^2_{min}(LO) - \chi^2_{min}(HO)|$. For LO, a $5$-year run of neutrinos has better sensitivity for $\delta_D > 0$ than for $\delta_D < 0$, and the opposite for a $5$-year run of antineutrinos. These are reversed for HO. Note that not every case has a large enough sensitivity. However, running with $2.5$ years each for neutrinos and antineutrinos can lead to stable sensitivity around $\Delta \chi^2 \approx 4$. These also apply to NO$\nu$A.\\[5mm]

Special attention should be paid to the equal running time scheme at NO$\nu$A and T2K. For the mass hierarchy and the octant of the atmospheric angle, an enhanced and stable sensitivity can be achieved in contrast to the single mode of running neutrinos or antineutrinos. Although it is a compromise in the sensitivity of distinguishing $\mathbb Z^s_2$ and $\overline{\mathbb Z}^s_2$, the sensitivity is still acceptable. With all these factors taken into consideration, it is better to adopt the equal running time scheme. Since our model with the residual $\mathbb Z^s_2$ or $\overline{\mathbb Z}^s_2$ symmetry is an example of the general case which is not constrained by any correlation such as (\ref{eq:cD}), the feature should also apply generally.

\section{Conclusion}
\label{sec:conclusion}

This paper explores the phenomenological consequences of the residual $\mathbb Z^s_2$ and $\overline{\mathbb Z}^s_2$ symmetries. The refined measurements on the reactor angle leads to distinct predictions of the Dirac {\tt CP} phase and the atmospheric angle between the two residual symmetries. For $\mathbb Z^s_2$, the predicted distribution of the {\tt CP} phase peaks around $\pm 60^\circ$, while the peak is around $\pm 140^\circ \sim \pm 145^\circ$ for $\overline{\mathbb Z}^s_2$. The Jarlskog invariant of the former is constrained within $0.02 < J_\nu < 0.04$, while it extends from $0$ to $0.04$ for the latter. The atmospheric angle obtains a broader distribution $30^\circ < \theta_a < 60^\circ$ from $\mathbb Z^s_2$ and $38^\circ < \theta_a < 52^\circ$ from $\overline{\mathbb Z}^s_2$, while the shape is controlled by the Dirac {\tt CP} phase. These show the possibility of precision neutrino experiments to distinguish between them. For accelerator type neutrino experiments, such as NO$\nu$A and T2K, the sensitivity on the mass hierarchy can reach $\Delta \chi^2 > 4$ for the region $0.42 \lesssim \sin^2 \theta_a \lesssim 0.7$ if $\mathbb Z^s_2$ is true and $0.43 \lesssim \sin^2 \theta_a \lesssim 0.58$ if $\overline{\mathbb Z}^s_2$ is true. With the mass hierarchy determined, the residual $\mathbb Z^s_2$ and $\overline{\mathbb Z}^s_2$ symmetries can be distinguished from each other if the {\tt CP} phase is within the quarter around $\delta_D = 0^\circ \mbox{ or } 180^\circ$ and $\mathbb Z^s_2$ is true. Within the other two quarters around $\delta_D = \pm 90^\circ$, both $\mathbb Z^2_s$ and $\overline{\mathbb Z}^s_2$ can be excluded if the atmospheric angle is not maximal. Otherwise, no sizable sensitivity can be achieved. All these results are obtained with a split schedule, a $1.5$-year run of neutrinos and a $1.5$-year run of antineutrinos at NO$\nu$A together with a $2.5$-year run of neutrinos and a $2.5$-year run of antineutrinos at T2K. This arrangement can significantly increase and stabilize the sensitivities to the mass hierarchy and the octant of the atmospheric angle with only a moderate compromise to the sensitivity of distinguishing $\mathbb Z^s_2$ and $\overline{\mathbb Z}^s_2$, in comparison to running with purely neutrinos or antineutrinos.

\section*{Acknowledgement}

We would like to thank Duane Dicus for his input and corrections in regards to this work. SFG is grateful to the University of Pittsburgh for the invitation to the neutrino workshop ``{\it Beyond $\theta_{13}$}" in February 2013, and to the Department of Physics and Astronomy at Michigan State University for their hospitality and a fruitful visit during which the current work was conceived. ADH and SFG greatly benefited from the discussion with Carl Bromberg about the NO$\nu$A experiment. This work is supported in part by Grant-in-Aid for Scientific research (No. 25400287) from JSPS. WWR was supported in part by the National Science Foundation under Grant PHY-1068020.

\section*{Notes added}

At the final stage of this work, there appeared two works \cite{Luhn1308,*Ballett,*Meloni1308} that also explore the phenomenological consequences of residual symmetries at neutrino experiments. 

\bibliographystyle{hunsrt}
\bibliography{z2pheno}
\nocite{*}

\end{document}